\documentclass[12pt]{article}
\usepackage{amssymb}
\usepackage{showlabels}

\begin{document}

\begin{titlepage}
\begin{center}
\begin{Large}
{\bf Unitary Evolution on the $\mathbb{Z}_{2^n}\times \mathbb{Z}_{2^n}$ Phase Space}

\end{Large}

\vskip3truecm

E. G. Floratos$^{(a)}$\footnote{E-mail:
mflorato@phys.uoa.gr} and
S. Nicolis$^{(b)}$\footnote{E-mail:
Stam.Nicolis@lmpt.univ-tours.fr. }

\vskip1truecm
$^{(a)}$ {\sl Physics Department, University of Athens, Athens, Greece}

\vskip0.5truecm

$^{(b)}$  {\sl CNRS--Laboratoire de Math\'ematiques et Physique
Th\'eorique (UMR 6083)\\
Universit\'e de Tours, Parc Grandmont, 37200 Tours, France}

\vskip5truecm

\begin{abstract}
We construct  quantum evolution operators on the space of states, 
that
 realize the metaplectic representation of the modular group
$SL(2,{\Bbb Z}_{2^n})$. 
This representation acts in a natural way on the
coordinates of the non-commutative 2-torus, ${\Bbb T}_{2^n}^2$ and thus is 
 relevant for noncommutative field theories as well as theories of
quantum space-time. The larger 
class of operators, thus defined, may be useful for the more efficient 
realization of new  quantum algorithms.

\end{abstract}
\end{center}

\end{titlepage}
Recent progress in M-theory indicates that spacetime itself becomes 
noncommutative at scales where D-branes play an important
role~\cite{matrixmodel,schwarz}.
This noncommutativity comes about in a rather natural way because
D-branes are charged, gravitational solitons, moving in backgrounds
with magnetic flux. This is reminiscent of the  Landau problem, where 
the noncommutativity of the two, real, space coordinates is brought
about by the magnetic flux\cite{connes}. The strength of the flux 
provides a measure of non-commutativity. 

Another, {\em a priori} independent,  manifestation of non-commutativity 
is realized in quantum mechanics. Here spacetime is commutative, but 
{\em phase space} is not. The ``strength'' of the non-commutativity is 
given by Planck's constant, $\hbar$. 

Non-commutativity in quantum mechanics implies that the phase space is
cellular: the size of the elementary cell, 
$\Delta \hat{q}\times \Delta\hat{p}$, is set by the quantum of action. 
It is not possible to realize   
$\hat{q}$ and $\hat{p}$ by finite dimensional matrices 
that respect the commutation relation,  
$[\hat{q},\hat{p}]={\mathrm i}I$. It is, however, well known how to 
overcome this obstruction: to pass from the algebra to the group, by 
taking exponentials, $P=\exp({\mathrm i}\hat{p})$, 
$Q=\exp({\mathrm i}\hat{q})$. This is only possible when $2\pi\hbar$ is a
rational number; we will study the simplest case, $2\pi\hbar=1/N$.
The operators $P$ and $Q$ satisfy $QP=\omega PQ$, where 
$\omega=\exp(2\pi{\mathrm i}/N)$ and admit an $N$--dimensional, irreducible, 
 representation\cite{Weyl}.
This discretization of  phase space thus differs markedly 
from the usual lattice regularization on spacetime. 
However, one can use this  to discretize spacetime in the presence of magnetic
flux of strength $1/N$ per cell. 

Classical mechanics in the Hamiltonian formalism amounts to the study 
of phase space symplectomorphisms. For one degree of freedom the linear 
symplectomorphisms  
form the group $SL(2,{\mathbb{R}})$ and the corrrespondance principle
allows us to find the unitary evolution operator $U({\sf A})$, that 
corresponds to the classical transformation ${\sf A}\in SL(2,{\mathbb{R}})$
\cite{cartier}. 
For the ${\mathbb{Z}}_N\times{\mathbb{Z}}_N$ phase space lattice, 
$SL(2,{\mathbb{R}})$ becomes $SL(2,{\mathbb{Z}}_N)$\cite{floratos1}.

In previous work~\cite{fqmdikamas,twotothenprel}
 we constructed the
metaplectic representation of $SL(2,{\Bbb Z}_N)$, which realizes Bohr's 
correspondance principle in a particularly transparent fashion. 
These quantum maps have been studied, in particular
within the context of quantum chaos~\cite{qchaos}, Rational Conformal
Field Theory~\cite{rcft} and quantum gravity~\cite{thooft}.  

The case $N=2^n$ was not amenable to analysis using the tools thus far 
available, although it is of clear interest for quantum
computing~\cite{qualgor} and the state space has been widely used in
communication engineering. The principal
difficulty resides in resolving ambiguities due to the 
factors of $1/2$ that abound in the expressions of the metaplectic
representation, indicative of theorems that hold for odd integers but not even 
ones!

These ambiguities are not just a technicality: they highlight  a 
fundamental difference between odd and even values of the discretization,
since even values allow the inclusion of fermions. As is well known, the
discretization of fermionic actions entails difficulties in taking into
account the quantization of classical symmetries. These {\em anomalies} become 
manifest as obstructions in the construction of local evolution amplitudes. 

In this note we shall present the construction of the evolution operator for 
the discretization $N=2^n$ of the phase space\footnote{A preliminary report may be
found in ref\cite{twotothenprel}--the present note supersedes that
presentation and completes that computation, 
using other techniques, and ties up loose ends tht had been left dangling there.} This will be possible for a 
subgroup of all possible classical symplectomorphisms, 
namely those that contain 
an {\em even} number of {\em odd} entries--and it is interesting to note that 
the Fourier transform belongs to this class!  
It is thus possible to put in
perspective the success of the Fourier transform in the development of quantum
information processing; but also to realize that it constitutes just a special
case, that is by no means singular.

Let us start by recalling the consistent quantization of linear
symplectomorphisms 
$$
\left(\begin{array}{c} r'\\ s'\end{array}\right)=\left({\sf A}\in 
SL(2,\mathbb{Z}_N)\right)
\left(\begin{array}{c} r\\ s\end{array}\right)
=\left(\begin{array}{cc} a & b \\ c & d\\ \end{array}\right)
\left(\begin{array}{c} r\\ s\end{array}\right)
$$
where $(r,s)$ and $(r',s')$ label the classical phase space and, along with
the elements  $,a,b,c,d$,
are integers mod $N$. 

It is possible to find, for every $N$,  operators $J_{r,s}$ that 
generate the Heisenberg--Weyl group and 
are given by 
$$
J_{r,s}=\omega^{r\cdot s/2}P^rQ^s
$$
where $QP=\omega PQ$ and $\omega=\exp(2\pi\mathrm{i}/N)$. 
In the basis where 
$$
Q_{k,l}=\omega^{k}\delta_{k,l}
$$
and 
$$
P_{k,l}=\delta_{k-1,l}
$$
the $J_{r,s}$ have matrix elements
$$
\left[J_{r,s}\right]_{k,l}=\delta_{k-r,l}\omega^{\frac{s}{2}(k+l)}
$$

We want to construct a unitary operator, $U({\sf A})$ that satisfies two requirements:

\begin{itemize}
\item It realize the {\em metaplectic} representation,
i.e. 
$$
U({\sf A})J_{r,s}U({\sf A})^{-1}=J_{(r,s){\sf A}}
$$
for all values of $k,l=0,\ldots,N-1$ and all values of the ``classical'' phase
space $(r,s)$. 

\item It realize a group representation: for any two ${\sf A}, {\sf B}\in 
SL(2,{\mathbb{Z}}_{2^n})$
$$
U({\sf A}\cdot {\sf B})=U({\sf A})\cdot U({\sf B})
$$
\end{itemize}
We shall start from the decomposition of the ``classical''matrix
\begin{equation}
\underbrace{\left(\begin{array}{cc} a & b \\ c & d\end{array}\right)}_{ {\sf
    A} }=
\underbrace{\left(\begin{array}{cc} 1 & bd^{-1} \\ 0 & 1\end{array}\right)}_{
  {\sf L} }
\underbrace{\left(\begin{array}{cc} d^{-1} & 0 \\ 0 & d\end{array}\right)}_{
  {\sf S} }
\underbrace{\left(\begin{array}{cc} 1 & 0 \\ cd^{-1} & 1\end{array}\right)}_{
  {\sf R} }
\end{equation}
and build up the operator $U({\sf A})$ from the corresponding ``building
blocks'' 
\begin{equation}
U\left[\left(\begin{array}{cc} a & b \\ c & d\end{array}\right)\right]=
U\left[\left(\begin{array}{cc} 1 & bd^{-1} \\ 0 & 1\end{array}\right)\right]
U\left[\left(\begin{array}{cc} d^{-1} & 0 \\ 0 & d\end{array}\right)\right]
U\left[\left(\begin{array}{cc} 1 & 0 \\ cd^{-1} & 1\end{array}\right)\right]
\end{equation}
which should satisfy the previous two requirements. 

We already note that this construction imposes the constraint that $d$ be an
odd number, in order that $d^{-1}\,\mathrm{mod}\,2^n$ exist. 

The classical motion described by the matrix
$$
\left(\begin{array}{cc} 1 & bd^{-1} \\ 0 & 1\end{array}\right)
$$
is a (right) translation in phase space with parameter $x=bd^{-1}$ and 
corresponds to the motion of a free particle. This leads us to write
\begin{equation}
U_{\mathrm{right}}(x)\equiv U\left[\left(\begin{array}{cc} 1 & x \\ 0 &
    1\end{array}\right)\right]_{k,l}=\omega^{Cxk^2}\delta_{k,l},\,\,k,l=0,\ldots,2^n-1
\end{equation}
with $C$ to be determined. 

It is easy to check that this {\em Ansatz} satisfies both properties under the 
condition that $C=-1/2$, which   is the only value compatible with the 
metaplectic property. 
This means that $x$ must be an even number, since $1/2\,\mathrm{mod}\,2^n$
doesn't exist. 

The translation opeator thus constructed has, however, a problem with
periodicity. 
$$
U\left[\left(\begin{array}{cc} 1 & x \\ 0 & 1\end{array}\right)\right]_{k,l}=
\delta_{k,l}\omega^{-xk^2/2}\Rightarrow
U\left[\left(\begin{array}{cc} 1 & x+2^n \\ 0 &
    1\end{array}\right)\right]_{k,l}=\omega^{-xk^2/2}\omega^{-2^{n-1}k^2}\delta_{k,l}\neq
U\left[\left(\begin{array}{cc} 1 & x \\ 0 & 1\end{array}\right)\right]_{k,l}
$$
which must be addressed, since the phase space is a torus. The remedy is to 
define the operator
\begin{equation}
M_{k,l}=(-1)^{k^2}\delta_{k,l}
\end{equation}
This operator has the property that 
\begin{equation}
\left[M^2\right]_{k,l}=\delta_{k,l}=I_{2^n\times 2^n}
\end{equation}
Therefore the operators 
\begin{equation}
C_\pm=\frac{I\pm M}{2}
\end{equation}
are {\em projectors} on the ``even'' and ``odd'' sublattices
respectively. 

The $M$ operator, furthermore, anticommutes with $P$:
$$
\begin{array}{l}
\left[M\cdot P\right]_{k,l}=(-1)^{k^2}\delta_{k,m}\delta_{m-1,l}=(-1)^{k^2}\delta_{k-1,l}\\
\left[P\cdot M\right]_{k,l}=\delta_{k-1,m}(-1)^{m^2}\delta_{m,l}=
(-1)^{l^2}\delta_{k-1,l}=-(-1)^{k^2}\delta_{k-1,l}=-\left[M\cdot
  P\right]_{k,l}
\end{array}
$$
This implies that $M$ {\em commutes} with $J_{2r,2s}$. If we thus define 
the ``twisted'' translation operators
\begin{equation}
T_{\pm}(x)=
U\left[\left(\begin{array}{cc} 1 & x \\ 0 &
    1\end{array}\right)\right]\cdot C_\pm
\end{equation}
then we may check that these also constitute a metaplectic representation on
the ``even'' sublattice, since
$$
J_{2r,2s}T_\pm(x)=T_\pm(x)J_{2r,2rx+2s}
$$
Furthermore, periodicity is, now, manifest:
$$
T_\pm(x+2^n)=U\left[\left(\begin{array}{cc} 1 & x+2^n \\ 0 &
    1\end{array}\right)\right]\cdot C_\pm=
U\left[\left(\begin{array}{cc} 1 & x \\ 0 &
    1\end{array}\right)\right]\cdot M\cdot C_\pm=
\pm U\left[\left(\begin{array}{cc} 1 & x \\ 0 &
    1\end{array}\right)\right]\cdot C_\pm
$$
since 
$$
M\cdot C_\pm=\pm C_\pm
$$
The (left) translation operator may be constructed from the right translation 
operator through the identity
$$
U\left[\left(\begin{array}{rr} 0 & 1 \\ -1 & 0\end{array}\right)
\left(\begin{array}{rr} 1 & -y \\ 0 & 1\end{array}\right)
\left(\begin{array}{rr} 0 & -1 \\ 1 & 0\end{array}\right)\right]=
U\left[\left(\begin{array}{rr} 0 & 1 \\ -1 & 0\end{array}\right)\right]
U\left[\left(\begin{array}{rr} 1 & -y \\ 0 & 1\end{array}\right)\right]
U\left[\left(\begin{array}{rr} 0 & -1 \\ 1 & 0\end{array}\right)\right]
$$
and the expression for the unitary operator that realizes the discrete Fourier
transform
$$
F\equiv U\left[\left(\begin{array}{rr} 0 & -1 \\ 1 & 0\end{array}\right)\right]_{k,l}=
\frac{1}{\sqrt{2^n}}\omega^{k\cdot l}
$$

The classical motion described by the diagonal matrix 
$$
\left(\begin{array}{cc} d^{-1} & 0 \\ 0 & d\end{array}\right)
$$
is a dilatation. We make the following {\em Ansatz} for the corresponding 
unitary operator 
\begin{equation}
U\left[
\left(\begin{array}{cc} d^{-1} & 0 \\ 0 & d\end{array}\right)\right]_{k,l}=
\delta_{k,d\cdot l}
\end{equation}
It is equally straightforward to verify that it satisfies the group property
and the metaplectic property. It is worth noting that this is a {\em
  permutation} operator and may be used to construct non-commutative 
solitons\cite{non-dispersive}. 

We thus obtain the matrix elements for the left translation operator 
\begin{equation}
\begin{array}{l}
\displaystyle U_\mathrm{left}(y)\equiv U\left[\left(\begin{array}{cc} 1 & 0 \\ y & 1\end{array}\right)\right]_{k,l}=\\
\displaystyle U\left[\left(\begin{array}{rr} 0 & 1 \\ -1 & 0\end{array}\right)\right]_{k,m}
U\left[\left(\begin{array}{rr} 1 & -y \\ 0 & 1\end{array}\right)\right]_{m,m'}
U\left[\left(\begin{array}{rr} 0 & -1 \\ 1 & 0\end{array}\right)\right]_{m',l}=\\
\displaystyle\frac{1}{\sqrt{2^n}}\omega^{-km}\omega^{ym^2/2}\delta_{m,m'}
\frac{1}{\sqrt{2^n}}\omega^{m'l}=
\frac{1}{2^n}\sum_{m=0}^{2^n-1}\omega^{ym^2/2+m(l-k)}
\end{array}
\end{equation}
This operator has, however, the same periodicity problem as the (na\"ive) right
translation operator! The remedy to this is to use $T_+(x)$ instead of 
$\omega^{-xk^2/2}\delta_{k,l}$ in the previous expression. This amounts to 
multiplying the above expression by the projector $(1+(-1)^m)/2$. The
expression for the left translation operator thus becomes
$$
U_{\mathrm{left}}(y)=F^{-1}T_+(-y)F
$$
and it obviously has the metaplectic property since all three of its factors
have it. It also has the correct periodicity properties, by virtue of the
presence  of the appropriate projector. 
It should, however, be stressed that the 
classical phase space is ``thinned'' out by a factor of 2, since the points 
are indexed as $(2r,2s)$ instead of simply $(r,s)$. 
 
The left translation operator thus obtained 
is a {\em generalized Gau\ss{ }sum}, 
\begin{equation}
\sigma_n(p,q)=\frac{1}{\sqrt{2^n}}\sum_{k=0}^{2^n-1}\omega^{pk^2+qk}
\end{equation}
that reduces to the usual Gau\ss{ } sum, $\sigma_n(p)$\cite{Lang}, for $q=0$.

The usual Gau\ss{ }sum, $\sigma_n(p)$, has the value
$$
\sigma_n(p)=(-2^n|p)\varepsilon(p)(1+\mathrm{i})
$$
where $(-2^n|p)$ is the {\em Jacobi symbol}\cite{Lang} (equal to 1 if $-2^n$ is 
a quadratic residue mod $p$ and $-1$ otherwise) and $\varepsilon(p)$ is equal to 1 if
$p\equiv\,1\,\mathrm{mod}\,4$ and $\mathrm{i}$ if
$p\equiv\,3\,\mathrm{mod}\,4$. The result is summarized in the following table:$$
\begin{array}{|c|c|l|}
\hline
p & q & \sigma_n(p,q) \\
\hline 
\mathrm{odd} & 0 & (-2^n|p)\varepsilon(p)(1+\mathrm{i})\\
\hline
2^mp' & 0 & \sqrt{2^m}(-2^{n-m}|p')\varepsilon(p')(1+\mathrm{i})\\
\hline
{\bf any} & \mathrm{odd} & 0\\
\hline
\mathrm{odd} & \mathrm{even} &
\omega^{-\frac{q^2}{4p}}(-2^n|p)\varepsilon(p)(1+\mathrm{i})\\
\hline
2^mp' & 2^lq' & 0\,\, (m\geq l)\\
\hline
2^mp' & 2^lq' & \sqrt{2^m}\omega^{-\frac{2^{2(l-m)-2}q'^2}{p'}}\sigma_{n-m}(p')\,\, (m<l)\\
 \hline
\end{array}
$$
Including the projector leads to computing the sum
$$
\begin{array}{l}
  \displaystyle
\frac{1}{\sqrt{2^n}}\sum_{k=0}^{2^n-1}\omega^{pk^2+qk}\left(\frac{1+(-1)^k}{2}\right)=
\frac{1}{\sqrt{2^n}}\sum_{k=2k'=0}^{2^n-1}\omega^{p(2k')^2+q(2k')}=\\
\displaystyle
\frac{1}{\sqrt{2^n}}\sum_{k'=0}^{2^{n-1}-1}\omega_{n-1}^{2p(k')^2+q(k')}=
\frac{1}{\sqrt{2}}\sigma_{n-1}(2p,q)
\end{array}
$$
We may now finish the calculation of the matrix elements for
the evolution operator, when $a$ and $d$ are odd and $b$ and $c$ are even:
\begin{equation}
\begin{array}{l}
\displaystyle
U\left[\left(\begin{array}{cc} a & b \\ c & d \end{array}\right)\right]_{k,l}=
\left[U_{\mathrm{right}}(b\cdot
d^{-1})U_{\mathrm{dilatation}}(d)U_{\mathrm{left}}(c\cdot
d^{-1})\right]_{k,l}=\\
\displaystyle
\omega^{bd^{-1}k^2/2}\left(\frac{1+(-1)^k}{2}\right)\frac{\sigma_{n-1}(cd^{-1},
kd^{-1}-l)}{\sqrt{2^{n+1}}}
\end{array}
\end{equation}

The case, when $a$ and $d$ are both even and $b$ and $c$ are both odd may be
related to this one through the observation
$$
\left(\begin{array}{cc}
\mathrm{even} & \mathrm{odd} \\
\mathrm{odd}  & \mathrm{even}\\
\end{array}\right)=
\left(\begin{array}{cc}
\mathrm{odd}' & \mathrm{even}' \\
\mathrm{even}'  & \mathrm{odd}'\\
\end{array}\right)
\left(\begin{array}{cc}
0  & -1 \\
1  & 0\\
\end{array}\right)
$$
Thus, if we {\em define}
\begin{equation}
U\left[\left(\begin{array}{cc}
\mathrm{even} & \mathrm{odd} \\
\mathrm{odd}  & \mathrm{even}\\
\end{array}\right)\right]=
U\left[\left(\begin{array}{cc}
\mathrm{odd}' & \mathrm{even}' \\
\mathrm{even}'  & \mathrm{odd}'\\
\end{array}\right)\right]
U\left[\left(\begin{array}{cc}
0  & -1 \\
1  & 0\\
\end{array}\right)\right]
\end{equation}
then we may deduce that this evolution operator has the metaplectic property,
since 
$$
\left.\begin{array}{l}
U({\sf A}_1{\sf A}_2)=U({\sf A}_1)U({\sf A}_2)\\
J_{r,s}U({\sf A}_j)=U({\sf A}_j)J_{(r,s){\sf A}_j}\,\,j=1,2\\
\end{array}\right\}\Rightarrow
J_{r,s}U({\sf A}_1{\sf A}_2)=U({\sf A}_1{\sf A}_2)J_{(r,s){\sf A}_1{\sf A}_2}
$$
The set of elements of $SL(2,{\mathbb{Z}}_{2^n})$ of the form
$$
\left(\begin{array}{cc}
\mathrm{odd} & \mathrm{even} \\
\mathrm{even}  & \mathrm{odd}\\
\end{array}\right)
$$
forms a normal subgroup. Together with the set of elements
$$
\left(\begin{array}{cc}
\mathrm{even} & \mathrm{odd} \\
\mathrm{odd}  & \mathrm{even}\\
\end{array}\right)
$$
it generates a bigger subgroup of $SL(2,{\mathbb{Z}}_{2^n})$, which we may
call $H$.  In this paper we have constructed the metaplectic representation of 
$H$ and have shown that it may be written 
in terms of the Fourier transform, the two translation operators and the
dilatation operator. 

On the other hand, classical symplectomorphisms of the form
$$
\left(\begin{array}{cc}\mathrm{odd} & \mathrm{odd}\\ \mathrm{odd} &
  \mathrm{even}\end{array}\right)
$$
cannot be quantized in the same way. It is intriguing that the cube of any
such element belongs to $H$ and {\em can} be consistently quantized by the 
procedure we have presented.

{\bf Acknowledgements}: S. N. would like to thank the theory group at the 
NRCPS ``Demokritos'' for the warm hospitality and M. Axenides for stimulating 
discussions. E. G. F. and S. N. also acknowledge the support and hospitality
of the CERN Theory Division where this work was completed. 

\end{document}